# Elastic models for the non-Arrhenius viscosity of glass-forming liquids

Jeppe C. Dyre, Tage Christensen, and Niels Boye Olsen,
Department of Mathematics and Physics (IMFUFA),
DG centre "Glass and time",
Roskilde University, Postbox 260,
DK-4000 Roskilde, Denmark.

**Abstract:**
*This paper first reviews the shoving model for the non-Arrhenius viscosity of viscous liquids. According to this model the main contribution to the activation energy of a flow event is the energy needed for molecules to shove aside the surrounding, an energy which is proportional to the instantaneous shear modulus of the liquid. Data are presented supporting the model. It is shown that the fractional Debye-Stokes-Einstein relation, which quantitatively expresses the frequently observed decoupling of, e.g., conductivity from viscous flow, may be understood within the model. The paper goes on to review several related explanations for the non-Arrhenius viscosity. Most of these are also "elastic models," i.e., they express the viscosity activation energy in terms of short-time elastic properties of the liquid. Finally, two alternative arguments for elastic models are given, a general solid-state defect argument and an Occam's razor type argument.*

## 1. Introduction

Glasses are formed from extremely viscous liquids. The properties of glasses are inherited from the liquid. Glass-forming liquids present a challenge from the basic science point of view because they exhibit universal features which are not well understood. The term "universality" – a favoured term in the vocabulary of physicists – refers to features which are shared by all glass-forming liquids whether they are oxide melts, polymers, molecular liquids, ionic liquids, metallic liquid alloys, or viscous liquids studied by computer simulations. The universal features relate to the temperature dependence of the viscosity and to the time dependence of relaxation processes. Here we shall only be concerned with the former. The central question is: Why is the viscosity of glass-forming liquids with few exceptions non-Arrhenius?

It would be easy to understand a viscosity with Arrhenius temperature dependence by reference to rate theory by arguing as follows: "A barrier of molecular origin is to be overcome in the flow processes, and the height of this barrier determines the activation energy of viscosity." There are, however, only few liquids that exhibit an Arrhenius viscosity (e.g., a pure silica melt); the vast majority of liquids have an activation energy which increases significantly upon cooling.

A liquid's (shear) relaxation time $\tau$ is related to its shear viscosity $\eta$ by Maxwell's famous expression

$$\tau = \frac{\eta}{G_\infty}, \qquad (1)$$

where $G_\infty$ is the instantaneous shear modulus of the liquid, i.e., the shear modulus measured on short time scales where the liquid does not have time to flow. Although $G_\infty$ varies with temperature, a fact which is crucial for the model discussed below, its temperature dependence is insignificant compared to the dramatic temperature dependencies of relaxation time and viscosity. Thus the two latter quantities are roughly proportional.

Well-known models for the non-Arrhenius behaviour are the free volume model[1,2,3] and the Adam-Gibbs entropy model.[4] In the free volume model, which exists in several versions, the idea is that the contraction of the liquid upon cooling strongly affects the rate of molecular motion because volume is needed for molecular rearrangements. The volume in excess of the molecular volume is termed the "free" volume. This quantity obviously decreases upon cooling, but it is not possible to define the free volume rigorously because a molecule does not have a definite volume.

The entropy model predicts that the activation energy of viscosity is inversely proportional to the configurational entropy. This quantity decreases upon cooling, thus giving rise to non-Arrhenius behaviour. The entropy model, which is generally regarded as experimentally vindicated, is based on several assumptions. One is the reasonable assumption that it is possible to split the entropy into two terms, configurational and vibrational entropy. Another assumption is that the size of a *cooperatively rearranging region* depends on the configurational entropy because the region must be large enough to contain at least two states (potential energy minima). This implies an increase in cooperativity upon cooling which is, however, not compelling. For instance, a crystal without dislocations flows via motion of vacancies or interstitials, and the flow process does not become more and more cooperative as temperature decreases - the number of defects just decreases. A further assumption which goes into the entropy model is the *ad hoc* assumption that the activation energy is proportional to the region volume.

When it comes to comparing the entropy model to experiments one may have concerns about some of the evidence reported in the literature.[5] Thus Johari[6] has critically examined the assumptions usually made in the experimental validation of the model, and the predicted identity of the Kauzmann temperature with the temperature, where the Vogel-Fulcher-Tammann expression for the viscosity diverges, was recently questioned by Tanaka.[7] Another problem is that the regions calculated from the model often contain only 3-4 molecules. This is inconsistent with the assumption of a region which is largely unaffected by its surroundings. Moreover, the configurational entropy is usually identified with the so-called excess entropy, i.e., liquid minus crystal entropy at the given temperature. The justification for this is ostensibly that the phonon (vibrational) contribution to the entropy is the same in liquid and crystal. Consequently, since a crystal has virtually zero configurational entropy, the vibrational entropy of the liquid is equal to the crystal entropy at any given temperature. This assumption is common, but it seems to be at variance with the following: The high-frequency sound velocity usually has significantly stronger temperature dependence in the viscous liquid than in the crystal. Since the vibrational entropy is determined by the phonon spectrum which also determines the high-frequency limit of the elastic constants and sound velocities, it seems that the vibrational entropy of the liquid cannot be identified with the crystal entropy.

Worries along the lines indicated above some time ago led us to re-examine the question of the origin of the non-Arrhenius viscosity. Thus an alternative model, the "shoving" model, was proposed in 1996.[8] In this paper we briefly summarize the reasoning behind the model and present new data supporting it. Related models are reviewed and two new arguments for the shoving model and similar models presented - a solid-state defect argument and an "Occam's razor" argument.

## 2. The shoving model

The starting point is a belief that viscous liquids are to be viewed more as "solids which flow" than as less-viscous liquids like ambient water. This appears to have been the views of both Kauzmann[9] and Goldstein.[10] The "solidity" point of view is justified as follows.[11] A glass-forming liquid close to the calorimetric glass transition has extremely large viscosity. Thus most molecular motion goes into vibrations, just like in a solid. Only rarely does anything happen in the form of a flow event, a molecular rearrangement. As Goldstein emphasized in his classical 1969 paper,[10] the barrier for a flow event must be considerably larger than the thermal energy $k_BT$ - this is why the viscosity is large. Thus a viscous liquid is most of the time indistinguishable from a (disordered) solid. This is confirmed by numerous computer simulations.

A forced solid also flows by *sudden*, *rare*, and *localized* molecular rearrangements - for a crystal in thermal equilibrium flow proceeds via motion of vacancies or interstitials, because there are no extended dislocations. The conclusion is that

$$\textit{Viscous liquid} \quad \cong \quad \textit{Solid which flows}.$$

The basic idea of the shoving model is the same as that of the free volume model, namely that *extra volume is needed* for a flow event to occur. The work done in creating this extra volume is identified with the activation energy. This is how the name of the model arose: In order to rearrange, the molecules must shove aside the surrounding molecules.

A flow event, like any barrier crossing, happens on a very fast time scale (picoseconds) by an unlikely thermal fluctuation. The fact that the fluctuation is *unlikely* accounts for the long time between two flow events (which in turn is the reason for the large viscosity). The fact that thermal fluctuations are *fast* means that the surrounding liquid behaves as a solid during a flow event. Thus standard solid-state elasticity theory may be used for calculating the "shoving" work.

The shoving model assumes the simplest possible flow event, one of spherical symmetry. Contrary to intuition, perhaps, the expansion of a sphere into a larger sphere in an elastic solid does not result in any compression of the surroundings. This fact comes from solving the equations of solid-state elasticity theory which show that the radial expansion varies with radius as $1/r^2$, thus inducing a zero-divergence displacement vector field (implying no density change). Consequently, the elastic constant determining the shoving work is the shear modulus. And since this all happens on a fast time scale, the shoving work is proportional to the *instantaneous* shear modulus, $G_\infty$.

In glass-forming liquids $G_\infty$ increases as temperature decreases, and $G_\infty$ is much more temperature dependent than, e.g., in crystals and glasses. We define the activation (free) energy $E$ from the viscosity relative to its high-temperature limit by $\ln(\eta/\eta_0) = E/k_B T$ (the same definition is used in the entropy model). The activation energy is not to be confused with the apparent activation energy, which is defined as the tangential slope in the Arrhenius plot. In general, the activation energy depends on temperature. A convenient measure of how much $E$ changes with temperature is the "index," $I = -d\ln E/d\ln T$,[12] which for glass-forming liquids typically is between 2 and 6. The index is defined at any temperature; its value at the glass transition determines the fragility, m, by the equation: $m = 16(1+I)$[12] where the number 16 is the logarithm of the ratio between the time scale defining the calorimetric glass transition (1000 s) and a typical molecular vibration time ($10^{-13}$ s).

The final expression for the temperature-dependent activation energy in the shoving model is

$$E(T) = V_C G_\infty(T),\qquad(2)$$

where $V_C$ is a "characteristic" microscopic volume which is assumed to be temperature independent.

To summarize, the basic assumptions behind Eq. (2) are:
1) The activation energy is [mainly] *elastic* energy.
2) This elastic energy is located in [mainly] the *surroundings* of the rearranging molecules.
3) The elastic energy is [mainly] *shear* elastic energy, i.e., not associated with any density change.

As shown elsewhere, if the contribution to the activation energy from the rearranging molecules decreases strongly as function of the volume of the energy barrier maximum, the main contribution to the activation energy is indeed mainly located in the surroundings. Thus the activation energy is mainly elastic energy. Moreover, although the elastic energy for a general non-spherical flow event also has a contribution from volume changes, even for highly anisotropic flow events this contribution is never more than 10% of the total elastic energy.[13]

The experiments reported below utilize a three-disc piezo-ceramic transducer making it possible to measure the shear modulus of highly viscous liquids at frequencies ranging from 1 mHz to 50 kHz.[14] Figure 1(a) gives the imaginary part of the frequency-dependent shear modulus divided by temperature for 5-polyphenyl-ether at different temperatures. Figure 1(b) presents the same plot for triphenyl phoshite. The dashed lines show that there is a linear relationship between maximum loss over temperature and the logarithm of the loss-peak frequency. The line extrapolates to zero at $10^{12}$ Hz, close to a typical attempt frequency. These data support the shoving model because the mechanical loss obeys time-temperature superposition for these two liquids (not shown); thus the maximum shear mechanical loss is proportional to the instantaneous shear modulus with a temperature-independent proportionality constant (this, of course, applies only if there are no further relaxation processes at frequencies higher than the ones probed). We conclude that the logarithm of the shear mechanical loss peak frequency is proportional to the instantaneous shear modulus over temperature, which is Eq. (2).

Other data supporting the model are given in Fig. 2. The figure shows viscosity as a function of temperature for several molecular liquids (full symbols) - clearly non-Arrhenius - and the same data plotted as function of the variable $X \propto G_\infty/T$ normalized to one at the glass transition (open symbols). There is good agreement with the model with a physically reasonable prefactor (left endpoint of the diagonal line). Some of these data refer to liquids also studied by Barlow and co-workers in 1967 by ultrasonics in the MHz region;[15] our measurements reproduce theirs, confirming the assumption of no significant losses above the highest frequency accessible by the piezo-ceramic shear transducer.

The shoving model may also be applied to structural relaxation. This was done in 1998 utilizing the Tool-Narayanaswami (TN) formalism with the reduced time defined by scaling with the relaxation time expression of the shoving model where, however, the instantaneous shear modulus changes with time as the liquid approaches equilibrium ("ages"). This version of the TN formalism was applied to describe relaxation of the instantaneous shear modulus itself of the DC704 silicone oil measured by means of the piezo-ceramic method in the resonance mode.[16] Recently, the shoving model was successfully applied to describe structural relaxation of bulk metallic glasses.[17]

A new application of the shoving model is that it provides a simple framework for quantitatively understanding the so-called decoupling between different relaxation processes often observed in viscous liquids.[18] Decoupling means that some relaxation processes are much faster than the dominant (alpha) process, the one with characteristic time related to viscosity via the Maxwell relation Eq. (1). The decoupling phenomenon was studied extensively in the 1990's and several models have been proposed which are able to explain it qualitatively and sometimes quantitatively.[19,20,21,22,23,24] As a typical example, ionic conductivity often decouples from viscosity in the liquid phase, a decoupling that makes ion conduction in the glasses practically possible. An intriguing empiricism referred to as the "fractional Debye-Stokes-Einstein relation" usually applies.[25] It states that upon temperature/pressure variations the conductivity $\sigma$ varies as $\sigma \propto \eta^{-C}$ where $C<1$ is a constant. Usually only temperature is varied, but recently Bielowka, Psurek, Ziolo, and Paluch in a study of two molecular liquids showed that the fractional Debye-Stokes-Einstein relation applies also when pressure is varied with the contant $C$ being pressure and temperature independent.[26] In terms of the activation energies of conductivity, $E_\sigma$, and viscosity, $E_\eta$, the Debye-Stokes-Einstein relation reads

$$E_\sigma(p,T) = CE_\eta(p,T). \tag{3}$$

How can this be understood? The decoupling phenomenon is hard to understand within the Adam-Gibbs entropy model where the basic concept is that of a cooperatively rearranging region. The shoving model, on the other hand, provides a possible explanation: It is reasonable to assume that there are only a finite number of types of flow event. Each type has an activation energy which is given by the shoving model expression with some characteristic volume. Of the different types of flow events a certain subset is needed for the system to "percolate" in the energy landscape, i.e., to obtain ergodicity. Among these, at low temperatures the flow event with the largest characteristic volume has the largest activation energy; this becomes the bottleneck which determines the activation energy of viscosity. For ion motion a parallel reasoning applies, except that the number of flow events involving ions is a subset of the total number of flow events. Dc conduction decouples from viscosity whenever the conducting ions need some, but not all of the flow events involved in viscous flow, in order to percolate. Of those needed for conduction the one with largest characteristic volume determines the dc conductivity activation energy. To summarize: Viscous flow has one characteristic volume, dc conduction has another, smaller characteristic volume. The ratio between these two is the constant C of Eq. (3). This constant is temperature and pressure independent, if all characteristic volumes are affected by pressure and temperature in the same way.

**3. Related "elastic" approaches for explaining the non-Arrhenius viscosity**

Several other models lead to the shoving expression Eq. (2) or to expressions which experimentally are hard to distinguish from it. We summarize these models in chronological order.

   a. The first reference to Eq. (2) seems to be that of a little known work[27] by Tobolsky, Powell, and Eyring from 1943. These authors discussed viscosity by adopting a harmonic approximation to the intermolecular potential. The idea was to "relate the curvature of the potential within which the atoms move to the elastic constants" whereby "the viscous flow process can be described by calculation of the rates with which molecules move from one equilibrium position to the next." Assuming a cosine potential for the molecule in question this led to Eq. (2). Tobolsky and co-workers did not discuss frequency dependence of the shear modulus, but clearly their "shearing modulus" must be the instantaneous shear modulus of the liquid. - If one assumes a constant Poisson ratio, shear and bulk moduli are proportional. As pointed out by Tobolsky et al., in this case the model predicts proportionality between the viscosity activation energy and the bulk modulus, which is consistent with a relationship postulated earlier by Gemant who showed that both quantities are proportional to the energy of vaporization.[28]
   b. A model related to the shoving model, but for a different phenomenon, is the 1954 Anderson-Stuart model for ion conduction in glass.[29] These authors argued that the barrier to be overcome for an ion jumping between two minima has two contributions: The electrostatic attraction between an ion and its neighbouring non-bridging oxygen atom and the elastic work needed for the ion to expand the structure during the jump. The latter contribution was estimated using solid-state elasticity theory, leading to Eq. (2). In the ionic case the Coulomb contribution usually cannot be ignored, though.

c. Mooney[30] in 1957 argued that "a liquid not only could be but perhaps should be treated as an elastic continuum with a stress relaxation mechanism." He proposed a theory according to which, whenever the local deformation in any region exceeds a certain critical value, the structure in the region loses its rigidity resulting in molecular rearrangements. From a rather intricate reasoning Mooney predicted that the viscosity activation energy is proportional to the square of the "thermal longitudinal" sound velocity. If this is denoted by $c$ and the mass of one molecule by $m$, Mooney's expression for the activation energy is

$$E \propto mc^2. \qquad (4)$$

Since sound velocities squared are proportional to mechanical moduli, Eq. (4) is equivalent to Eq. (2) if the shear and bulk moduli are proportional (in their temperature and pressure variations).

d. The model of Bueche from 1959[31] also relates the relaxation time to elastic properties of the surroundings, now based on an argument involving concentric shells of surrounding molecules. The idea is that, "if these shells vibrate outward in phase, the innermost shell would expand greatly, leaving the central molecule in a rather large hole so it could move to a new position." Bueche's calculation of the probability of this happening leads to an expression that differs somewhat from Eqs. (2) and (4), but the physical picture is obviously close to that suggested by Mooney.

e. Nemilov in 1968 arrived at Eq. (2) from a completely different line of reasoning.[32] He noted that for several glasses there is proportionality between the glass shear modulus and the glass transition temperature. To explain this he combined Eyring's expression for the viscosity prefactor with Dushmann's expression for the rate prefactor (barrier energy divided by Planck's constant), leading via the Maxwell relation to Eq. (2). Equation (2) explains the proportionality between the glass transition temperature and the glass shear modulus, because the glass shear modulus is equal to the instantaneous shear modulus of the liquid at the glass transition temperature.

f. Hall and Wolynes[33] in 1987 reasoned along lines similar to those of Tobolsky, Powell, and Eyring, by basing their argument in the simplest version on a harmonic approximation around energy minima. This led to the prediction that the logarithm of the viscosity (or average relaxation time) is proportional to $1/<x^2>$, where $<x^2>$ is the mean-square displacement of the molecules in their vibrations around a potential energy minimum. We shall refer to this prediction as the "harmonic approximation:"

$$\frac{E}{k_B T} \propto \frac{a^2}{<x^2>} . \qquad (5)$$

Here $a$ is the intermolecular distance which is usually regarded as constant, while the vibrational mean-square displacement often depends on temperature as well as on pressure. A similar reasoning was applied long ago by Flynn[34] for the calculation of activation energies of point defect mobilities in metals. - Of course, at high pressure the intermolecular distance does change; in this case Eq. (5) may be expressed in terms of the density $\rho$ as follows

$$\frac{E}{k_B T} \propto \frac{\rho^{-2/3}}{<x^2>}. \qquad (6)$$

After Hall and Wolynes many authors discussed and used the harmonic approximation for describing the dynamics of glass-forming liquids (see, e.g., the references of Ref. 12); recently Eq. (5) was connected to the free volume model by Starr, Sastry, Douglas, and Glotzer, who confirmed it by computer simulations.[35] The crucial point is that in viscous liquids the vibrational mean-squared displacement is generally quite temperature dependent, often increasing much more rapidly with temperature than in a perfectly elastic solid where it is proportional to temperature.

g. In the approximation where all contributions to vibrations come from longitudinal and transverse phonons, the harmonic approximation Eq. (5) is equivalent to Eq. (2) if the shear modulus is replaced[12] by a combination of shear and (isothermal) longitudinal instantaneous moduli. It was recently shown, that the *temperature dependence* of the relevant combination of these elastic constants is dominated by that of the shear modulus since at most 8% comes from the bulk modulus.

h. Ngai recently[36] argued that the coupling model also implies the harmonic approximation.

This survey shows that the idea of connecting fast and slow dynamics - vibrations and viscous flow - has been around almost as long as glass science has been an academic discipline. At first it may seem puzzling that dynamics on time scales which differ 12-15 orders of magnitude could be connected. However, while the long time scale of viscous flow is determined by the time *between* two flow events, the short time scale (the picosecond time scale) is more or less that of the barrier transition *itself*. In view of this, properties on the short time scale may very well be relevant for determining the actual barrier height and thus "setting the clock" for the long time scale.

Finally, it should be mentioned that recently two most interesting correlations between fast and slow dynamics were reported. To appreciate these, recall that glass properties correspond to fast properties of the liquid just above the glass transition. Scopigno, Roucco, Sette, and Monaco[37] showed that the vibrational properties of the glass well below the glass transition correlate with the fragility of the liquid, while Novikov and Sokolov[38] showed that the degree of non-Arrhenius behaviour of the liquid correlates with the Poisson ratio of the glass.[39] How these intriguing findings may relate to the above discussed models is not clear at present.

**4. A solid-state defect argument**

The physical picture of a viscous liquid as a "*solid which flows*" is based on the fact that on time scales much shorter than the Maxwell relaxation time virtually no molecules have moved away from their initial position, where they simply vibrate just as in a solid. Thus, e.g., a viscous liquid with Maxwell relaxation time equal to 1 second simulated on the fastest computer available today is indistinguishable from a solid, albeit a non-crystalline solid. A flow event takes the molecules of the "solid" from one potential energy minimum to another, just like an activated process in a genuine solid, and just like in a solid these events are extremely rare.

The "$liquid \cong solid$" viewpoint invites one to comparing viscous liquid flow events to defect motion in crystals. In equilibrium a crystal contains no extended dislocations, so virtually all defects are point defects. In an interesting book by Varotsos and Alexopoulos (VA) from 1986[40] the main message is that point defect creation free energies - *as well as* the free energies of activation for self-diffusion - quite generally are proportional to the isothermal bulk modulus (with proportionality constant equal to some microscopic volume). If it is assumed that bulk and shear moduli are proportional in their temperature/pressure variations, the VA prediction becomes equivalent to the shoving model prediction Eq. (2). This shows that the shoving model and related models - most of which, as we have seen, are based on the harmonic approximation in some form – should be viewed in a more general setting than so far has been done: Viscous flow is like other diffusion processes in solids, the only new ingredient is that the relevant elastic constants are the high-frequency moduli which are generally much more temperature dependent in viscous liquids than in crystals or glasses.

A simple way to understand why point defect energies are proportional to mechanical moduli is by reference to the "strain-energy model," where a defect is regarded as a distortion of a continuum having the properties of the macroscopic crystal. This model was proposed by Zener in 1942 to account for the decrease of density of heavily cold-worked metals.[41] The model was later shown to explain the activation volumes derived from pressure effects on solid-state diffusion.[42] It is the same reasoning that is used in the shoving model, except that here it is used for calculating a barrier height, not a defect creation energy.

## 5. An Occam's razor argument

Several authors have compared Eq. (4) or equivalent formulations in terms of Einstein or Debye temperatures or mechanical moduli to experiment on the viscosity of glass-forming liquids. A straightforward way to make this comparison is to utilize the fact that the liquid relaxation time has a definite value at the glass transition. Thus the glass transition temperature is proportional to the activation energy of the liquid at this temperature with a universal proportionality constant, and the glass transition temperature may be estimated by probing properties *of the glass* (moduli, sound velocities, etc), which vary relatively little below the glass transition.

Generally, there is good agreement. Some authors use the longitudinal sound velocity of the glass, sometimes the transverse sound velocity is used. Examples of comparisons to experiment for various classes of glass-forming liquids include Nemilov's 1968 paper, a paper from 1994 by Heuer and Spiess,[43] or more recent works by Sanditov, Sangadiev and Kozlov[44] and by Wang, Wen, Zhao, Pan and Wang.[45] The last paper is notable because it confirmed Eq. (4) for bulk metallic glasses.

Heuer and Spiess[43] presented an interesting argument for the glass transition temperature being proportional to $mc^2$. Inspired by the standard Lennard-Jones potential often used for computer simulation of simple liquids, which has just one energy scale, they argued that there is *only one relevant energy scale* also for real liquids. This energy scale is identified in the combination of molecular mass and sound velocity of Eq. (4). Comparing to data for oxide glass formers, polymers, molecular liquids, ionic solutions, and chalcogenide glasses, Heuer and Spiess concluded that $k_B T_g \propto mc^2$ is always obeyed, which indeed *must* be the case if there is only one relevant energy scale (for a more or less universal intermolecular potential). This implies Eq. (4) at $T_g$ because, as mentioned, the activation energy at the glass transition is a universal constant times $k_B T_g$.

We now extended the Heuer-Spiess line of reasoning into an "Occam's razor" argument for the shoving model / harmonic approximation. First, note that the Heuer-Spiess focus on the possibly of just one relevant energy, as emphasized by the authors, is consistent with the Lindemann melting criterion and the well-known empirical 2/3 rule which gives the ratio between glass transition temperature and melting temperature $T_m$: The Lindemann melting criterion states that melting of a crystal takes place when the vibrational displacement is 10% of the interatomic distance; this translates into a proportionality between $k_B T_m$ and $mc^2$ where $c$ is the crystal sound velocity. Assuming that the crystal and glass sound velocities are roughly equal and only weakly temperature dependent, the Lindemann melting criterion and the 2/3 rule *imply* Eq. (4) for the activation energy at the glass transition, where $c$ is now the glass sound velocity.

Turning the argument on its head, the 2/3 rule and the Lindemann melting criterion taken together strongly indicate that effectively just one energy scale is relevant for viscous liquids – otherwise, it is hard to understand why the glass transition temperature is never much smaller or much larger than the melting temperature. Returning to the problem of understanding the origin of the non-Arrhenius viscosity, we note that by this argument the activation energy *must* be given by Eq. (4). A logical consequence is that, if the liquid really has just one relevant energy scale, one would expect an almost Arrhenius viscosity. Real viscous liquids, however, are much more complex than Lennard-Jones liquids, and one expects "renormalization" of the effective energy scale, in other word the effective energy scale may very well depend on temperature. In this spirit, the simplest would be that Eq. (4) applies not only *at* the glass transition but in the whole range of temperatures *above* it as well. This is where Occam's razor enters. Finally, since the glass sound velocity is identical to the liquid high-frequency sound velocity at the glass transition, in order for Eq. (4) to apply in the liquid phase $c$ is to be identified with a high-frequency sound velocity (and not for the low-frequency sound velocity).

If $c$ is replaced by the transverse high-frequency sound velocity of the liquid, Eq. (4) as already mentioned is equivalent to the shoving model. If, however, $c$ is the high-frequency longitudinal sound velocity, a somewhat different prediction is arrived at which gives an activation energy that generally does not vary enough with temperature to account for the non-Arrhenius viscosity. In view of the models discussed above this is because the longitudinal sound velocity involves the *adiabatic* longitudinal modulus, not the isothermal, and thus does not properly reflect the mean-square displacement of Eq. (5) (of course, for the shear deformations the adiabatic and isothermal moduli are equal).

## 6. Concluding remarks

Based on the above overview we would like to suggest that the shoving model and several related models should be referred to collectively as "elastic models." The common idea of elastic models is that flow events in a viscous liquid are similar to point defect motion in solids, and that the non-Arrhenius viscosity of glass-forming liquids is caused by temperature variations of the short-time elastic constants.

To many researchers glass science is attractive in large parts because of the beauty of the entropy model and the intriguing possibility of an infinite relaxation time at the Kauzmann temperature. This beauty is lost if elastic models prevail. On the other hand, there may be important practical consequences if elastic models are correct. Thus the activation energy may be monitored directly by measuring the instantaneous shear modulus (via a piezo-ceramic transducer, standing wave ultrasonic techniques, ordinary or stimulated transverse Brillouin scattering, etc), and this opens up a new and direct way of monitoring and thus optimizing annealing processes below the glass transition. It is straightforward to adopt the Tool-Narayanaswami formalism with a reduced time definition based on the Eq. (2) expression for the activation energy (which itself changes with time during annealing), and it would certainly be a novel development if one, as predicted by the elastic models, were able to directly monitor the reduced time.

*Acknowledgments:*

The authors are indebted to Kristine Niss for constructive critiques of the manuscript. This work was supported by a grant from Danmarks Grundforskningsfond (DG) - the Danish National Research Foundation - for funding the centre for viscous liquid dynamics "Glass and Time".

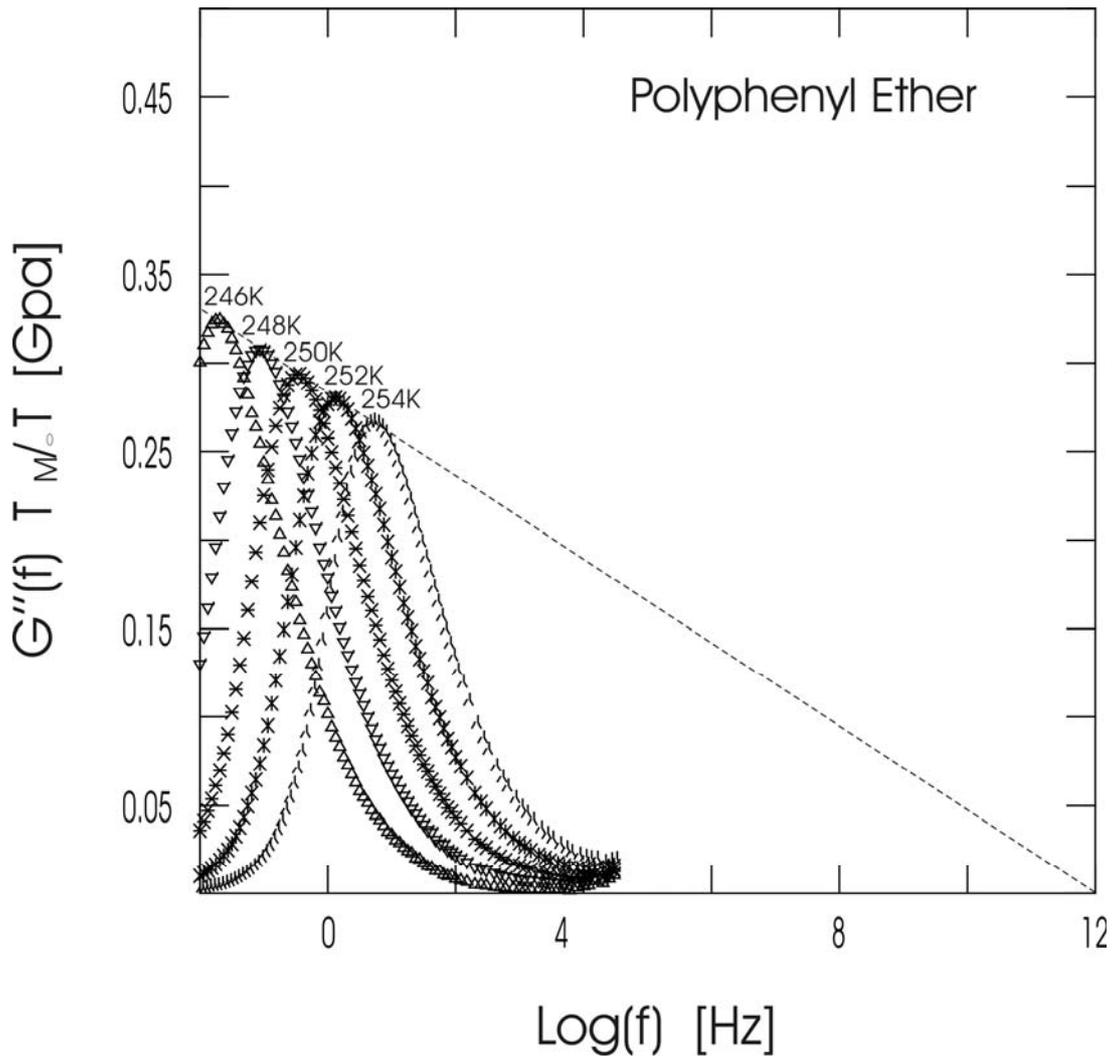

**Figure 1(a):**

Imaginary part of the frequency-dependent shear modulus divided by temperature normalized by the highest measured temperature, $T_M = 254$ K, for 5-polyphenyl-ether measured by the piezo-ceramic transducer. This liquid obeys time-temperature superposition for the shear modulus (not shown), a fact which implies that the maximum value of the loss is proportional to the instantaneous shear modulus. The shoving model thus predicts that the maxima extrapolate linearly to zero at a physically reasonable "attack" frequency (identified with the attempt frequency).

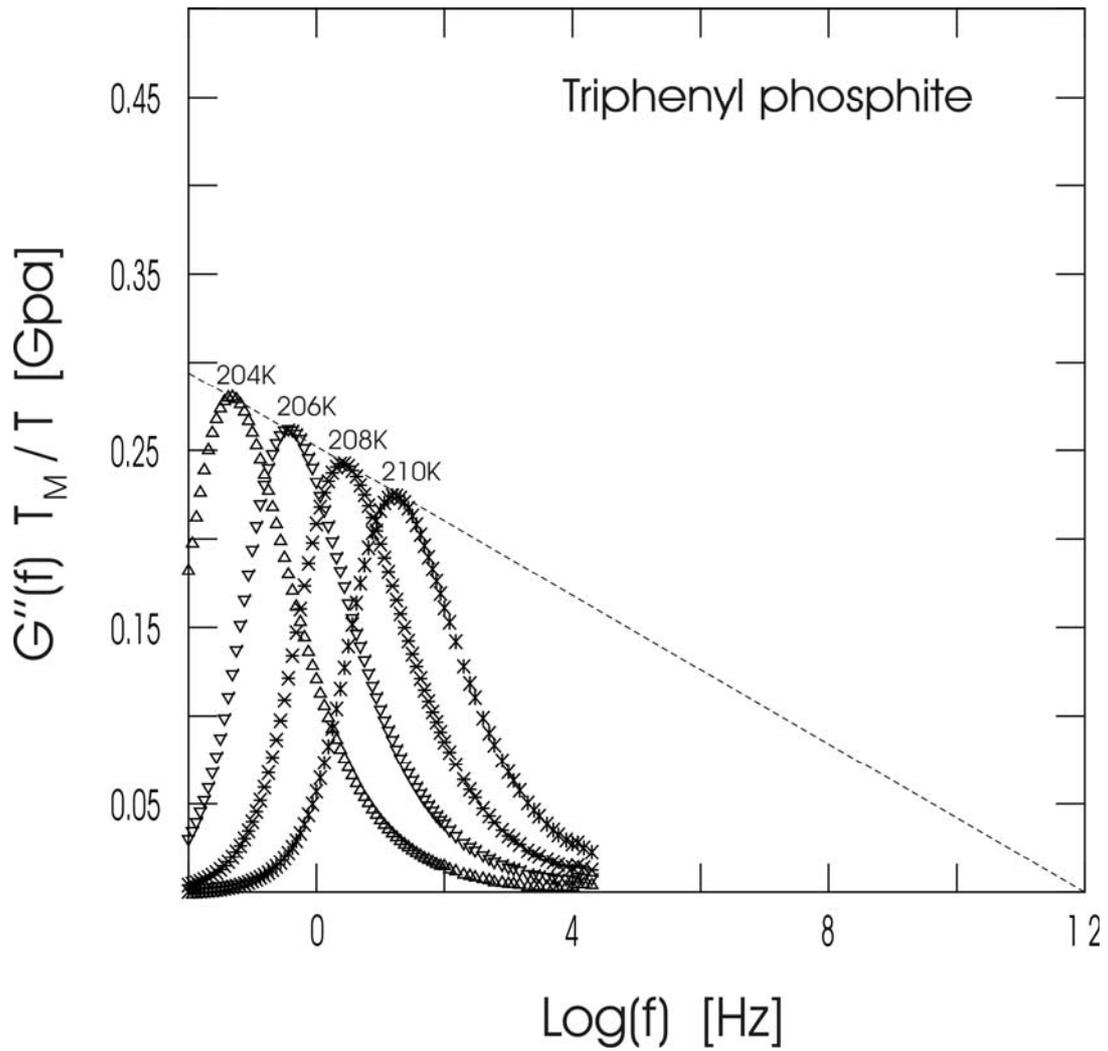

**Figure 1(b):**

Imaginary part of the frequency-dependent shear modulus divided by temperature normalized by the highest measured temperature, $T_M = 210$ K, for triphenyl phosphite measured by the piezo-ceramic transducer. This liquid also obeys time-temperature superposition for the shear modulus.

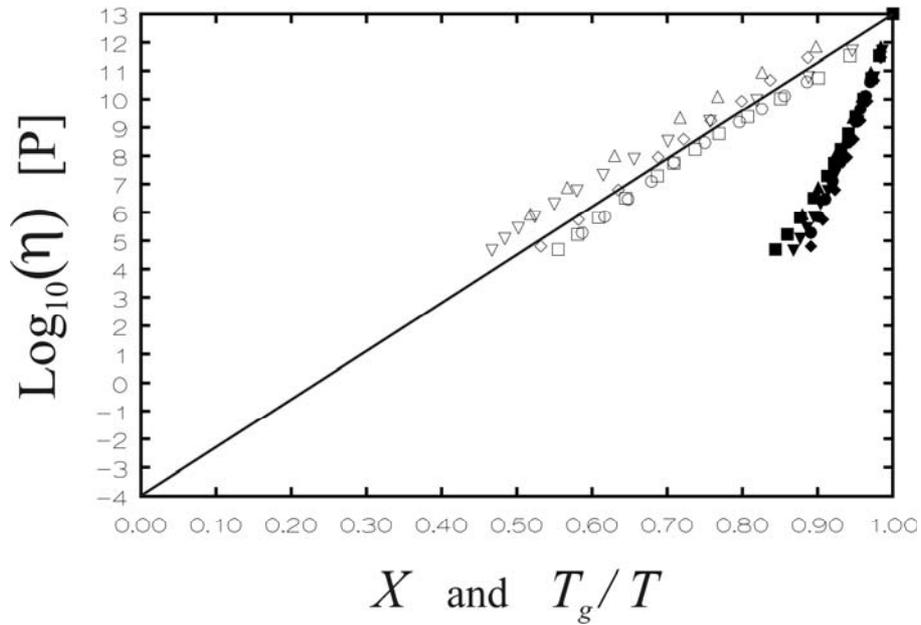

**Figure 2:**

Data first published in Ref. 8 showing the viscosity for 5 molecular liquids (4-methylpentan-2-ol, dioctyl phthalate, phenyl salicylate, dibutyl phthalate, tetra-methyl-tetra-phenyl-trisiloxane) as function of inverse temperature (full symbols) and as function of the variable $X$ which is defined as the quantity $G_\infty(T)/T$ normalized to one at the glass transition temperature (open symbols), where the latter is defined as the temperature where the equilibrium liquid viscosity is $10^{13}$ Poise ($10^{12}$ Pa s). The shoving model predicts that the viscosity as function of $X$ lies on the diagonal line starting at the lower left corner, corresponding to a typical high-temperature viscosity and thus a physically reasonable prefactor.